\documentclass[10pt,journal,compsoc]{IEEEtran}

\newtheorem{definition}{Definition}

\usepackage{multirow}
\usepackage{amsmath}

\usepackage{algpseudocode}
\usepackage[ruled,linesnumbered]{algorithm2e}

\ifCLASSOPTIONcompsoc
  \usepackage[nocompress]{cite}
\else
  \usepackage{cite}
\fi

\ifCLASSINFOpdf
   \usepackage[pdftex]{graphicx}

\else
\fi

\usepackage{array}

  \usepackage[caption=false,font=footnotesize,labelfont=sf,textfont=sf]{subfig}
  \usepackage[caption=false,font=footnotesize]{subfig}

\newcommand\MYhyperrefoptions{bookmarks=true,bookmarksnumbered=true,
	pdfpagemode={UseOutlines},plainpages=false,pdfpagelabels=true,
	colorlinks=true,linkcolor={blue},citecolor={blue},urlcolor={blue},
	pdftitle={Towards Target Sequential Rules},
	pdfsubject={Typesetting},
	pdfauthor={Wensheng Gan},
	pdfkeywords={Data mining, rule mining, targeted-oriented querying, target sequential rule.}}
\ifCLASSINFOpdf
\usepackage[\MYhyperrefoptions,pdftex]{hyperref}
\else
\usepackage[\MYhyperrefoptions,breaklinks=true,dvips]{hyperref}
\usepackage{breakurl}
\fi

\usepackage{url}

\begin{document}
\title{Towards Target Sequential Rules}

\author{Wensheng Gan, Gengsen Huang, Jian Weng, Tianlong Gu, Philip S. Yu,~\IEEEmembership{Fellow,~IEEE}
	
	\IEEEcompsocitemizethanks{
		
	\IEEEcompsocthanksitem Wensheng Gan is with the College of Cyber Security, Jinan University, Guangzhou 510632, China; and with Pazhou Lab, Guangzhou 510335, China.  (E-mail: wsgan001@gmail.com) 
	
	\IEEEcompsocthanksitem Gengsen Huang, Jian Weng, and Tianlong Gu are with the College of Cyber Security, Jinan University, Guangzhou 510632, China. (E-mail: hgengsen@gmail.com, cryptjweng@gmail.com, and gutianlong@jnu.edu.cn)

	\IEEEcompsocthanksitem Philip S. Yu is with the University of Illinois at Chicago, Chicago, USA. (E-mail: psyu@uic.edu)} %
	
	\thanks{Corresponding author: Wensheng Gan}
}

\IEEEtitleabstractindextext{%
\begin{abstract}
	
In many real-world applications, sequential rule mining (SRM) can provide prediction and recommendation functions for a variety of services. It is an important technique of pattern mining to discover all valuable rules that belong to high-frequency and high-confidence sequential rules. Although several algorithms of SRM are proposed to solve various practical problems, there are no studies on target sequential rules. Targeted sequential rule mining aims at mining the interesting sequential rules that users focus on, thus avoiding the generation of other invalid and unnecessary rules. This approach can further improve the efficiency of users in analysing rules and reduce the consumption of data resources. In this paper, we provide the relevant definitions of target sequential rule and formulate the problem of targeted sequential rule mining. Furthermore, we propose an efficient algorithm, called targeted sequential rule mining (TaSRM). Several pruning strategies and an optimization are introduced to improve the efficiency of TaSRM. Finally, a large number of experiments are conducted on different benchmarks, and we analyze the results in terms of their running time, memory consumption, and scalability, as well as query cases with different query rules. It is shown that the novel algorithm TaSRM and its variants can achieve better experimental performance compared to the existing baseline algorithm. 

\end{abstract}

\begin{IEEEkeywords}
	Data mining, rule mining, targeted-oriented querying, target sequential rule.
\end{IEEEkeywords}}

\maketitle

\IEEEdisplaynontitleabstractindextext

\IEEEpeerreviewmaketitle

\section{Introduction} \label{sec:introduction}

\IEEEPARstart{S}{equential} pattern mining (SPM) \cite{agrawal1995mining, mooney2013sequential,okolica2020sequence} is a popular technique in data discovery, which is widely used in various real-life application scenarios \cite{luna2019frequent,fournier2017survey}. The task of SPM is to discover all frequent sequential patterns from the sequence database, and traditional SPM algorithms are based on the frequency metric. The frequency of a frequent sequential pattern should be greater than the user-defined minimum support (\textit{minsup}). In other words, the number of occurrences of the frequent sequential pattern in the database reaches the minimum number required by users. Those discovered sequential patterns possess higher value in pattern analysis because they take into account the chronological order of events that appear in the database. For example, users usually buy a bike in a store first and then buy an inflator after a while. In this case, we can use the sequential pattern $s$ $=$ $<$\textit{bike}, \textit{inflator}$>$ for the representation. With this sequential pattern $s$, it is clear that the order of purchase of these two items can be known. If we ignore the chronological order in the pattern, the purchase order of these two items will become blurred. Reviewing some previous work, there are a number of applications that utilize related SPM algorithms, such as analysis of user purchasing behavior \cite{srikant1996mining}, analysis of DNA sequences \cite{floratou2011efficient}, and analysis of user web page click events \cite{fournier2012using}. To date, there are many algorithms \cite{mooney2013sequential, fournier2017survey,gan2019survey} of SPM designed to address specific requirements. In recent years, with the upgrading of hardware, related researchers are more interested in mining more valuable and compact sequential patterns in specific applications than in improving the efficiency of the algorithms. As far as we know, there are several major types of algorithms that researchers have focused on. Such as closed SPM \cite{yan2003clospan, wang2007frequent, gomariz2013clasp, fumarola2016clofast}, maximal SPM \cite{guan2005mining, garcia2006new, fournier2013mining, fournier2014vmsp}, and nonoverlapping SPM \cite{wu2017nosep, wu2020netncsp}. Those algorithms can reduce the number of patterns in the generation by imposing restrictions on the sequential pattern format, thus obtaining more valuable sequential patterns. Although there are many fast, efficient, and concrete algorithms of SPM in use today, the main drawbacks of these algorithms still exist. A sequential pattern that considers the temporal order cannot provide a probability to predict the next possible event or item. This means that SPM is very limited in scenarios such as recommended applications and predictive services. For example, in a weather forecast, if we convert the most recent events into these following sequences $t_1$ $=$ $<$\textit{dark clouds}, \textit{air humidity}, \textit{rain tomorrow}$>$ and $t_2$ $=$ $<$\textit{dark clouds}, \textit{air dry}, \textit{rain tomorrow}$>$. Obviously, the third event of both sequences is "rain tomorrow". If we observe the first two events of these two sequences, they are very different. According to the real phenomena, it is more likely to rain when the dark clouds are dense and the air is humid. However, we cannot know the probability of rain from these two sequences. If we analyze only these two sequences, we may mistakenly think that dark cloud is the biggest cause of rain. And whether the air is humid or not does not have much influence on whether it rains tomorrow.

Considering the main drawbacks of the algorithms of SPM, sequential rule mining (SRM) \cite{zaki2001spade, lo2009non, fournier2012cmrules, fournier2014erminer, fournier2015mining} incorporates the concept of confidence to solve these issues. SRM aims to discover a complete set of sequential rules from the sequence database that satisfy the user-defined minimum support (\textit{minsup}) and minimum confidence (\textit{minconf}). This allows the sequential rules not to focus only on frequency as a metric, which can serve recommendation scenarios \cite{jannach2014recommendation} and prediction applications \cite{das1998rule}. Reanalyzing the example discussed above, two sequential rules $r_1$ $=$ $<$\textit{dark clouds}, \textit{air humidity}$>$ $\rightarrow$ $<$\textit{rain}$>$ and $r_2$ $=$ $<$\textit{dark clouds}, \textit{air dry}$>$ $\rightarrow$ $<$\textit{rain tomorrow}$>$ are found by SRM. Where the confidence of the first rule is 90\%, and the second rule is 10\%. Then we can learn that whether the air is humid or not also has an effect on whether it rains tomorrow or not. The former rule states that when the sky with dense dark clouds and the air is humid, there is a 90\% chance that it will rain tomorrow. Although SRM can mine valuable and interesting sequential rules, it still has a bunch of problems. One of the main problems is that a large number of sequential rules will be generated if \textit{minsup} or \textit{minconf} is tuned down or when processing large datasets.

There is no doubt that as the size of the processed dataset increases or the related thresholds of the parameter are set small, the number of sequential rules that are mined increases a lot. If we focus on the sequential rules in which the target event occurs among them, then there is no need for other unrelated sequential rules. Given these two sequential rules $r_1$ $=$ $<$\textit{dark clouds}, \textit{air humidity}$>$ $\rightarrow$ $<$\textit{rain tomorrow}$>$ and $r_2$ $=$ $<$\textit{white clouds}, \textit{air dry}$>$ $\rightarrow$ $<$\textit{sunny tomorrow}$>$ in a weather forecast, if we only focus on the rules that affect whether it will be sunny tomorrow, then the rule $r_1$ is useless and is not necessary to be tapped. These unrelated rules are not helpful for users to perform analysis. In addition, if we want to know the effect of air humidity on tomorrow's weather, then we should get the antecedent of the sequential rules containing "air humidity", for example $t_1$ $=$ $<$\textit{dark clouds}, \textit{air humidity}$>$ $\rightarrow$ $<$\textit{rain tomorrow}$>$ and $t_2$ $=$ $<$\textit{dark clouds}, \textit{air humidity}$>$ $\rightarrow$ $<$\textit{cloudy tomorrow}$>$ are two useful rules. However, there is no current research on target sequential rules. The existing studies on target rules are based on association rules \cite{kubat2003itemset, lavergne2012min}, while association rules do not take into account the chronological order of events or items in the database. Although the general SRM algorithms can be used as naive approaches to discover the target sequential rules, they are simply too inefficient.

In order to solve the above-mentioned problems, we formally define the concept of target sequential rule and the framework of targeted sequential rule mining in this paper. We are the first work to perform the mining of target sequential rules. To improve the efficiency of the generation of target sequential rules, we propose several pruning strategies including unpromising transactions pruning strategy (UTP), unpromising items pruning strategy (UIP), unpromising 1 * 1 rules pruning strategy (URP), and unpromising extended items pruning strategy (UEIP). In addition, a hash table-based optimization is designed. A novel and efficient algorithm called TaSRM is proposed. The main contributions of our paper are as follows.

\begin{itemize}
	\item  To the best of our knowledge, there is no previous work on target sequential rules, and we present a total set of definitions of target sequential rules. Besides, we fully discuss the query cases of targeted sequential rule mining and formulate the problem of targeted sequential rule mining.
	
	\item  We propose an efficient TaSRM algorithm and its variants to discover a complete set of all target sequential rules for a query rule. Meanwhile, the number of target sequential rules obtained by TaSRM is absolutely correct. We perform this verification using a naive approach based on the SRM algorithm.
	
	\item  To reduce the running time and memory consumption of the mining process, we design four pruning strategies, including UTP, UIP, URP, and UEIP. We also design a post-processing technique for the traditional SRM algorithms, allowing them as naive methods to discover correct target sequential rules. Moreover, two hash tables are introduced as an optimization for TaSRM.
	
	\item  We conduct extensive experiments using different real and synthetic datasets and analyze the efficiency of the TaSRM algorithm in terms of running time, memory consumption, count of expansions, and different query cases. The efficiency and effectiveness of the TaSRM algorithm is shown from comprehensive aspects.
\end{itemize}

The remainder of this paper is organized as follows. We review the related work in Section \ref{sec:relatedwork}. The important definitions and the problem of targeted sequential rule mining are presented in Section \ref{sec:preliminary}. And then, we propose our TaSRM algorithm and several pruning strategies in Section \ref{sec:algorithm}. Subsequently, the experimental results are given and evaluated in Section \ref{sec:experiments}. Finally, we conclude our work and discuss the future studies in Section \ref{sec:conclusion}.

\section{Related Work}  \label{sec:relatedwork}

In this section, we discuss the main algorithms of sequential rule mining and targeted pattern querying.

\subsection{Sequential Rule Mining}  \label{section:SRM}

Sequential rule mining (SRM) \cite{zaki2001spade, lo2009non, fournier2012cmrules, fournier2014erminer, fournier2015mining} was proposed to solve the deficiency that the patterns mined by sequential pattern mining (SPM) \cite{agrawal1995mining, mooney2013sequential, okolica2020sequence} do not have confidence. SRM compensates for the limitations of SPM in the recommendation and prediction. Sequential rules can carry more favorable knowledge for deduction. Compared with association rule mining (ARM) \cite{agrawal1994fast, pasquier1999discovering, geng2006interestingness}, the rules obtained by SRM have the same merit as SPM, i.e., both consider the temporal relationship between items. Unlike the sequential pattern, a sequential rule can be represented as $X$ $\rightarrow$ $Y$, and $X$ $\cap$ $Y$ $=$ $\emptyset$. For this rule, $X$ is its antecedent, and $Y$ is its consequent. Each sequential rule has a support value and a confidence value. A mined sequential rule is qualified and frequent if its support is greater than the user-defined minimum support (\textit{minsup}) and its confidence is greater than the user-defined minimum confidence (\textit{minconf}). The order of items in $X$ or $Y$ can be ordered or unordered. According to this criterion, sequential rule mining can be divided into two types. One is partially-ordered SRM, another is totally-ordered SRM. Regardless of that type of SRM, for a sequence containing this rule, any item of $X$ appears before $Y$. Since the rules obtained from partially-ordered SRM are more valuable, we mainly discuss the studies of partially-ordered SRM.

RuleGen \cite{zaki2001spade} is the first algorithm to discover all sequential rules from a sequence database. RuleGen generates rules by comparing two sequential patterns of different lengths. As the sequential patterns are mined, the sequential rules are generated one after another. This method is obviously very inefficient. And then, Lo \textit{et al.} \cite{lo2009non} explored the relevant theoretical knowledge and proposed an algorithm called CNR for mining non-redundant sequential rules. Based on the composition of two pattern sets, the concept of non-redundant sequential rule was formally defined. In order to improve the efficiency, an algorithm called  MNSR-Pretree \cite{pham2014efficient} using attributed prefix-trees was proposed. The prefix-tree structure and an efficient pruning strategy can allow its performance much better. In addition to the totally-ordered SRM, partially-ordered SRM is a more valuable direction. CMDeo \cite{fournier2012cmrules} and CMRule \cite{fournier2012cmrules} were proposed to accelerate the mining process of sequential rules. CMRule is a two-phased-based algorithm using an association rule algorithm. It first utilizes an association rule algorithm to generate association rules and then generates sequential rules from association rules. This leads to the efficiency of CMRules being limited by the ARM algorithms. CMDeo first introduced the idea of left-expansion and right-expansion. The SRM algorithm using these two expansions is also a single-phase-based algorithm. In the initial stage, CMDeo generates all 1 * 1 sequential rules and then grows these rules recursively. To further improve the efficiency of CMDeo, an algorithm based on the idea of PrefixSpan \cite{han2001prefixspan} called RuleGrowth \cite{fournier2015mining} was proposed. RuleGrowth is a rule-growth-based algorithm that uses the occurrence flags to grow sequential rules. It can avoid generating those sequential rules that not appear in the database and also can work on the multiple-items-based sequences. Another fast algorithm is ERMiner \cite{fournier2014erminer}, which uses the method of merging equivalence classes to generate sequential rules. A data structure, namely sparse counting matrix, is designed to reduce the unnecessary rule generation. Although ERMiner runs in much less time than RuleGrowth, it consumes a lot of memory. In addition, a number of extensions to SRM have been proposed. By combining the concept of utility, related algorithms \cite{huang2021us, zida2015efficient} of high-utility sequential rule mining can work on utility-driven tasks. TNS \cite{fournier2013tns} is an efficient algorithm that can find out top-$k$ non-redundant sequential rules. TRuleGrowth \cite{fournier2015mining} was proposed to find sequential rules appearing with a window constraint. Furthermore, DUOS \cite{gan2021anomaly} was proposed to detect all utility-aware outlier sequential rules on network security and anomaly detection.

\subsection{Targeted Pattern Querying}
\label{section:TPQ} 

Targeted sequential pattern mining can find out all patterns that users interest and reduce the output of unrelated patterns. This improves the efficiency of pattern mining and reduces resource consumption, including time and space, etc. For targeted pattern mining, an important question is what pattern is the target and is more conspicuous? To date, there are some algorithms proposed to solve the problem of target pattern queries. As far as we know, these algorithms mainly address the case of target itemset and target association rule queries. In an earlier work, Kubat \textit{et al.} \cite{kubat2003itemset} proposed a data structure called an Itemset Tree. With the Itemset Tree, their query system can handle three different query cases. The first one is the pattern support query, the second one is the target itemset query, and the last one is the target association rule query. And then, Min-Max Itemset Tree \cite{lavergne2012min} was proposed to deal with dense and categorical datasets. It solves the two main problems of Itemset-Tree and reduces the number of searches for the nodes of subtrees. Since Itemset Tree uses a top-down search method, Fournier-Viger \textit{et al.} \cite{fournier2013meit} subjected the Itemset Tree to node compression and proposed MEIT. The leaner nodes allow MEIT to save up to 60\% of memory usage. In order to address the two issues of Itemset Tree that not being able to utilize the Apriori property during the query process and the generated rules being restricted, Lewis \textit{et al.} \cite{lewis2019enhancing} improved the Itemset Tree to improve the performance of Itemset Tree and provide a redesigned itemset generation process. Recently, a model called query-constraint-based ARM (QARM) \cite{abeysinghe2017query} was introduced to solve the query problem for clinical datasets. It is a query algorithm with constraints. GFP-Growth \cite{shabtay2021guided} was proposed for the scenario of multitude-targeted itemsets query in big data, and it is based on the idea of tree. In the utility mining task, TargetUM \cite{miao2021targetum}  is the first algorithm in terms of querying with the high-utility target itemset. It utilizes a lexicographic querying tree and some pruning strategies to quickly get query responses.

In addition to algorithms for itemset queries and association rule queries, there are also some algorithms proposed to query the target sequential patterns. Chiang \textit{et al.} \cite{chiang2003goal} focused on those target sequential patterns that the last itemset of those patterns is the query itemset and introduced the definition of goal-oriented sequential pattern. Chueh \cite{chueh2010mining}  proposed the method of filtering the database and reversing the database to discover all target sequential patterns that satisfy the time-intervals constraint. A PrefixSpan-based algorithm \cite{chand2012target} that combine the RFM model was proposed to find out all valuable old customers. Huang \textit{et al.} \cite{huang2022taspm} redefined the concept of the target sequential pattern and proposed a generic framework to discover interesting target patterns on sequences. Besides, by further considering the definition of the target sequential pattern and combining the utility values, TUSQ \cite{zhang2021tusq} was proposed to solve the query problem in utility-driven mining.

\section{Preliminary and Problem Statement} 
\label{sec:preliminary}

In this section, to better illustrate the targeted sequential rule mining, we briefly introduce the concepts and definitions used in the previous work. Based on these concepts and definitions, we formalize the problem of targeted sequential rule mining.

\subsection{Preliminary}

\begin{definition}[Sequence database]
	\rm Let $I$ = \{$i_{1}$, $i_{2}$, $\cdots$, $i_{n}$\} be an infinite set of $n$ distinct items. Each item can be considered as a symbolic representation of a real object. A set containing multiple different items is called itemset, which is denoted as $X$ and satisfies $X$ $\subseteq$ $I$. We assume that all items in an itemset are sorted in alphabetical order, i.e. $a$ $\textless$ $b$ $\textless$ $c$ $\textless$ $\cdots$. The alphabetical order is denoted as $\succ_{lex}$. A sequence is a list containing multiple itemsets, which is denoted as $S$. A sequence is also a transaction record, and the order of each itemset in it is the occurred order. For a sequence database $\mathcal{D}$, it is a collection containing multiple sequences and each sequence owns a unique identifier. The size of a sequence database $\mathcal{D}$ is the total number of sequences it contains, and can be denoted as $\vert$$\mathcal{D}$$\vert$. 
\end{definition}

For example, as shown in Table \ref{table1}, this sequence database has seven items and five sequences. The identifiers of these sequences are $s_1$, $s_2$, $s_3$, $s_4$, and $s_5$ respectively. We use this example database to illustrate the definitions and examples that follow.

\begin{table}[!htbp]
	\centering
	\caption{Sequence database}
	\label{table1}
	\begin{tabular}{|c|c|}  
		\hline 
		\textbf{sid} & \textbf{Sequence} \\
		\hline  
		\(s_{1}\) & $<$($b$), ($c$, $e$)$>$ \\ 
		\hline
		\(s_{2}\) & $<$($b$), ($a$), ($d$), ($f$)$>$ \\  
		\hline  
		\(s_{3}\) & $<$($b$, $c$), ($d$), ($a$), ($g$), ($b$, $c$)$>$ \\
		\hline  
		\(s_{4}\) & $<$($b$), ($a$), ($d$), ($e$), ($c$, $d$)$>$ \\
		\hline
		\(s_{5}\) & $<$($c$, $d$), ($a$, $b$), ($e$, $g$), ($c$)$>$ \\
		\hline
	\end{tabular}
\end{table}

\begin{definition}[Sequential rule]
	\label{definition:SR}
	\rm In this paper, we focus on the partially-ordered sequential rule. A partially-ordered sequential rule can be denoted as $r$ $=$ \{$X$\} $\rightarrow$ \{$Y$\}, $X$ $\neq$ $\emptyset$, $Y$ $\neq$ $\emptyset$, and $X$ $\cap$ $Y$ $=$ $\emptyset$. $X$ is called the antecedent of this rule, and $Y$ is called the consequent. In a sequence containing $r$, every one of the items of $X$ appears before $Y$. If this sequence contains $n$ itemsets and there is existing one integer $k$ (1 $\le$ $k$ $\textless$ $n$) that satisfies $X$ $\subseteq$ $\bigcup_{i=1}^{k}$ $I_i$ and $Y$ $\subseteq$ $\bigcup_{i=k+1}^{n}$ $I_i$. There is no overlap itemset between the itemsets containing $X$ and the itemsets containing $Y$. In this case, we also call $r$ appearing in this sequence. The size of the sequential rule $r$ is presented as $m$ * $f$, where $m$ is the size of $X$, and $f$ is the size of $Y$. Given another sequential rule $t$ with the size $h$ * $l$, if $h$ $\ge$ $m$ and $l$ $\textgreater$ $f$; or $h$ $\textgreater$ $m$ and $l$ $\ge$ $f$ hold, we can say that $t$ is larger than $r$.
\end{definition}

For example, in Table \ref{table1}, given a sequential rule $r$ $=$ \{$a$, $b$\} $\rightarrow$ \{$d$\}, its size is 2 * 1. This rule appears in $s_2$ and $s_4$. In these two sequences, we can see that both the occurred order of $a$ and $b$ is before $d$. Given another sequential rule $t$ = \{$c$, $d$\} $\rightarrow$ \{$a$, $b$\}, its size is 2 * 2. Sequences $s_3$ and $s_5$ contain this rule, and $t$ is larger than $r$.

\begin{definition}[The support and confidence of a sequential rule]
	\rm We use \textit{sids}($r$) to denote the set of identifiers for sequences in the database $\mathcal{D}$ that contain this sequential rule $r$ $=$ \{$X$\} $\rightarrow$ \{$Y$\}. \textit{sup}($X$) denotes the total number of sequences in the database $\mathcal{D}$ containing $X$. Therefore, the support of the rule $r$ can be defined as \textit{sup}($r$) $=$ $\vert$\textit{sids}($r$)$\vert$, and the confidence of it can be defined as \textit{conf}($r$) $=$ \textit{sup}($r$) / \textit{sup}($X$). In order to find all frequent and high-confidence sequential rules, the user needs to provide the minimum support (\textit{minsup}) and the minimum confidence (\textit{minconf}).
\end{definition}

For example, in Table \ref{table1}, given a sequential rule $r$ $=$ \{$b$, $d$\} $\rightarrow$ \{$g$\}. In database $\mathcal{D}$, \textit{sids}($r$) $=$ \{$s_3$, $s_5$\}, and \textit{sup}($r$) $=$ 2. Since there are four sequences containing the itemset ($b$, $d$), then the confidence of $r$ can be calculated as \textit{conf}($r$) $=$ 2 / 4 = 0.5.

\begin{definition}[Target sequential rule]
	\label{definition:TSR}
	\rm For targeted sequential rule mining, an important question is what sequential rules are the target rules? Given a sequential rule for a query, we call it a query rule, denoted as \textit{qr} $=$ \{\textit{XQuery}\} $\rightarrow$ \{\textit{YQuery}\}. For this query rule, there are three cases to consider, representing the three possible queries. The first one is that \textit{XQuery} is empty, i.e., querying which sequential rules can be derived with the consequent containing \textit{YQuery}. The second case is that \textit{YQuery} is empty, i.e. what sequential rules can be deduced by querying the antecedent with \textit{XQuery}. The last case is not empty for \textit{XQuery} and \textit{YQuery}. To accommodate these three cases, we specify that for a query rule \textit{qr} $=$ \{\textit{XQuery}\} $\rightarrow$ \{\textit{YQuery}\}, all of its target sequential rules \textit{tr} $=$ \{$X$\} $\rightarrow$ \{$Y$\} satisfy that \textit{XQuery} $\subseteq$ $X$ and \textit{YQuery} $\subseteq$ $Y$. When both  \textit{XQuery} and \textit{YQuery} are empty sets, this means that all sequential rules are target sequential rules. At that time, the mining process is also the traditional sequential rule mining, and we call \textit{qr} an empty query rule.
\end{definition}

\subsection{Problem Statement}
\label{Problem statement}

Based on the above definitions, we define the following problem. Given a sequence database $\mathcal{D}$, a query rule \textit{qr} $=$ \{\textit{XQuery}\} $\rightarrow$ \{\textit{YQuery}\}, a user-defined \textit{minsup}, and a user-defined \textit{minconf}. The problem of targeted sequential rule mining is to find out all target sequential rules \textit{tr} of \textit{qr} that satisfy \textit{sup}(\textit{tr}) $\ge$ \textit{minsup} and \textit{conf}(\textit{tr}) $\ge$ \textit{minconf}.

For example, in Table \ref{table1}, given a query rule \textit{qr} $=$ \{$a$, $b$\} $\rightarrow$ \{$c$\}, \textit{minsup} $=$ 2, and \textit{minconf} = 0.6. All target sequential rules of this query rule \textit{qr} in the example database $\mathcal{D}$ are shown in Table \ref{table2}. Even though other sequential rules are high-support and high-confidence, they are not the target rules. When we set \textit{qr} to an empty query rule, 40 rules are generated. Obviously, targeted sequential rule mining can reduce a large amount of unrelated sequential rules, thus avoiding unnecessary resource overhead.

\begin{table}[!htbp]
	\centering
	\caption{All target sequential rules of \textit{qr} $=$ \{$a$, $b$\} $\rightarrow$ \{$c$\}}
	\label{table2}
	\begin{tabular}{|c|c|c|c|}  
		\hline 
		\textbf{Id} & \textbf{Targeted sequential rule} & \textbf{Support} & \textbf{Confidence}\\
		\hline  
		\(r_{1}\) & \{$a$, $b$\} $\rightarrow$ \{$c$\} & 3 & 0.75 \\ 
		\hline
		\(r_{2}\) & \{$a$, $b$, $d$\} $\rightarrow$ \{$c$\} & 3 & 0.75 \\  
		\hline  
		\(r_{3}\) & \{$a$, $b$, $d$, $e$\} $\rightarrow$ \{$c$\} & 2 & 1.0\\
		\hline  
		\(r_{4}\) & \{$a$, $b$, $d$, $g$\} $\rightarrow$ \{$c$\} & 2 & 1.0\\
		\hline
		\(r_{5}\) & \{$a$, $b$, $e$\} $\rightarrow$ \{$c$\} & 2 & 1.0 \\
		\hline
		\(r_{6}\) & \{$a$, $b$, $g$\} $\rightarrow$ \{$c$\} & 2 & 1.0  \\
		\hline
	\end{tabular}
\end{table}

\section{The TaSRM Algorithm} \label{sec:algorithm}

In this section, we first provide some relevant definitions and discussions about targeted sequential rule mining, and then propose our pruning strategies and an optimization. Finally, we show our TaSRM algorithm.

\subsection{Definitions and Pruning Strategies}

\begin{definition}[Expansion]
	\rm Expansion is an important operation of the rule-growth-based algorithms \cite{fournier2012cmrules, fournier2015mining}. We also use this operation in this paper. For a sequential rule $r$ $=$ \{$X$\} $\rightarrow$ \{$Y$\} with the size $m$ * $n$, it can perform a left-expansion to generate a ($m$ + 1) * $n$ rule, or a right-expansion to generate a $m$ * ($n$ + 1) rule. For an item $i$ belonging to $I$, after performing a left-expansion, the expanded sequential rule is $r^\prime$ $=$ \{$X$\} $\cup$ $i$ $\rightarrow$ \{$Y$\}. And after performing a right-expansion, the expanded sequential rule is $r^\prime$ $=$ \{$X$\} $\rightarrow$ \{$Y$\} $\cup$ $i$. To avoid redundant sequential rules to be generated, we assume that the expanded item must follow the $\succ_{lex}$ order and that once the left-expansion is performed, the right-expansion cannot be performed. This regulation is not unique, and the relevant explanation can be referred to Refs. \cite{fournier2012cmrules, fournier2015mining}.
\end{definition}

For example, in Table \ref{table1}, for a sequential rule $r$ $=$ \{$b$\} $\rightarrow$ \{$g$\}, we can know that the item in \{$c$, $d$, $e$, $f$\} can be performed a left-expansion and no item can be performed a right-expansion. This is because $g$ is the largest item in order $\succ_{lex}$. If $r$ performs a left-expansion on $f$, the sequential rule $r^\prime$ $=$ \{$b$, $f$\} $\rightarrow$ \{$g$\} can be obtained.

\begin{definition}[Rule occurrence]
	\rm As we define the concept for a sequential rule $r$ in Definition \ref{definition:SR}, there is no overlap itemset between the itemsets containing the antecedent of $r$ and the itemsets containing the consequent of $r$ and there exists a required order. To find out which items can be expanded, TaSRM scans the sequences containing the sequential rule $r$. To speed up the scanning process, for each sequential rule, a binary list is used to save two positions on a sequence containing it. The binary list contains two variables, one of which is \textit{firstItemset} and the other is \textit{lastItemset}. We specify that the position subscript of the sequence starts from 0. \textit{firstItemset} denotes the earliest position that the subsequence from the first itemset to the \textit{firstItemset}-th itemset in the sequence is able to contain the antecedent of the sequential rule $r$ completely. Therefore, we should scan the sequence starting from \{\textit{firstItemset} + 1\}-th itemset to find items that can be performed the right-expansion. Backward scanning the sequence, \textit{lastItemset} denotes the earliest position that the subsequence from the \textit{lastItemset}-th itemset to the last itemset in the sequence is able to contain the consequent of $r$ completely. Hence, the positions of the items that can be performed the left-expansion in the sequence are before \textit{lastItemset}.
\end{definition}

For example, in Table \ref{table1}, for a sequential rule $r$ $=$ \{$b$\} $\rightarrow$ \{$g$\} and in $s_5$, the \textit{firstItemset} and the \textit{lastItemset} of this rule are 1 and 2 respectively. We should find the right-expandable item after subscript position 1 and the left-expandable item before subscript position 2 of the sequence. Consider another sequential rule $t$ $=$ \{$b$\} $\rightarrow$ \{$d$\} in $s_4$, the \textit{firstItemset} and the \textit{lastItemset} of this rule are 0 and 4 respectively, instead of 0 and 2.

\begin{definition}[Post-processing technique]
	\rm Since we first propose the framework of targeted sequential rule mining, there are no relevant algorithms to discover target sequential rules for a query rule. We use this post-processing technique, allowing traditional sequential rule mining algorithms as naive approaches to save all mined sequential rules belonging to target sequential rules for a query rule. In the rule generation, if a high-support and high-confidence sequential rule is not a target sequential rule of the query rule, it can be filtered. With this technique, the sequential rule mining algorithms can get all target sequential rules and filter out unrelated sequential rules. Obviously, this technique is very clumsy and inefficient. It is equivalent to selecting the rules that users are interested in from all the obtained sequential rules.
\end{definition}

\textbf{Strategy 1: Unpromising transaction pruning strategy (UTP)}. We use the unpromising transactions pruning strategy (UTP) to filter some transaction records  from the database that are unrelated and do not affect the targeted sequential rule mining. For a query rule \textit{qr} $=$ \{\textit{XQuery}\} $\rightarrow$ \{\textit{YQuery}\}, there are some cases that need to be considered. The first case is a transaction record that contains \textit{YQuery} but not \textit{XQuery}. In this case, we can safely filter out this transaction record. This is because this transaction record is unlikely to contain any target sequential rule for \textit{qr} and also not affects the calculation of support of other target rules. The second case is a transaction record that contains \textit{XQuery} but not \textit{YQuery}. In this case, this transaction record cannot be filtered. It is because if we filter this transaction record, then the calculation of confidence of the target sequential rules for \textit{qr} will be wrong. The third case we should consider is that a transaction record does not contain \textit{XQuery} and \textit{YQuery}. It is clear that we can safely filter this transaction record. This transaction record does not affect the mining. The last case is a transaction record that both contains \textit{XQuery} and \textit{YQuery}. Obviously, this transaction record cannot be filtered, otherwise the mining result is incorrect. We can summarize as follows. If a transaction record does not contain the \textit{XQuery}, then it can be safely filtered from the database.

For example, in Table \ref{table1}, for a query rule \textit{qr} $=$ \{$a$, $b$\} $\rightarrow$ \{$c$\}, we only need to filter the sequence $s_1$ and do not need to filter the sequence $s_2$. If we filter $s_2$, \textit{tr} $=$ \{$a$, $b$\} $\rightarrow$ \{$c$\} as one of the target sequential rules of \textit{qr} is outputted, while its confidence is equal to 1.0. Obviously, the confidence of \textit{tr} is wrong.

\begin{definition}[Transaction matching position map, \textit{TMPM}]
	\rm To facilitate the use of the pruning strategies proposed later, we design the data structure \textit{TMPM}. For each sequence, the corresponding two variables are saved in \textit{TMPM}. The first variable is called \textit{leftEnd} and the second variable is called \textit{rightEnd}. Forward scanning a transaction record, \textit{leftEnd} is used to record the position where the transaction record first exactly matches \textit{XQuery}. Backward scanning a transaction record, \textit{rightEnd} is used to record the position of the first exactly matching \textit{YQuery} of the transaction record. For the filtered database, each transaction record definitely contains \textit{XQuery}, but partially does not contain \textit{YQuery}. If a transaction record does not contain \textit{YQuery}, set both its \textit{leftEnd} and \textit{rightEnd} to the last position of the transaction record. This transaction record is only used to calculate the associated confidence values.
\end{definition}

For example, in Table \ref{table1}, for a query rule \textit{qr} $=$ \{$a$, $b$\} $\rightarrow$ \{$c$\}, we can know that the \textit{leftEnd} and \textit{rightEnd} of $s_2$ are both 3. It is because $s_2$ does not contain the \textit{YQuery} of \textit{qr}. While in $s_3$, the \textit{leftEnd} and \textit{rightEnd} of $s_3$ are 2 and 4, respectively.

\textbf{Strategy 2: unpromising items pruning strategy (UIP)}. We use the unpromising items pruning strategy (UIP) to filter out some unpromising items from the database, and UIP depends on \textit{TMPM}. In a transaction record, we first remove the items that appear in \textit{XQuery} and their positions are after the \textit{leftEnd} of this transaction record. This is because these items already appear before \textit{leftEnd}. Second, we remove the items that appear in \textit{YQuery} and their positions are before the \textit{rightEnd} of the transaction record. Similarly, these items already appear after \textit{rightEnd}. Finally, we calculate the support of the items in the database and remove the infrequent items.

For example, in Table \ref{table1}, for a query rule \textit{qr} $=$ \{$a$, $b$\} $\rightarrow$ \{$c$\}, we can know that after using UIP, $s_2$ becomes $<$($b$), ($a$), ($d$)$>$, $s_3$ becomes $<$($b$), ($d$), ($a$), ($g$), ($c$)$>$, and $s_5$ becomes $<$($d$), ($a$, $b$), ($e$, $g$), ($c$)$>$.

\textbf{Strategy 3: unpromising 1 * 1 rules pruning strategy (URP)}. We use \textit{firstLeftI} and \textit{firstRightI} to denote the smallest items in \textit{XQuery} and \textit{YQuery}. For any two items $m$ and $n$ belonging to $I$, the sequential rule \{$m$\} $\rightarrow$ \{$n$\} is not generated if $m$ is greater than \textit{firstLeftI} or $n$ is greater than \textit{firstRightI}. This is because we specify that the extended item in the rule growth of a sequential rule $r$ must be larger than the items in the antecedent or the consequent of $r$.

For example, for a query rule \textit{qr} $=$ \{$a$, $b$\} $\rightarrow$ \{$c$\}, we can know that \textit{firstLeftI} is $a$ and \textit{firstRightI} is $c$. Therefore, \{$d$\} $\rightarrow$ \{$c$\} will not be generated, even if it is a high-support and high-confidence sequential rule. 

\begin{definition}[\textit{XQuery} matching position and \textit{YQuery} matching position]
	\rm To better utilize the unpromising extended items pruning strategy (UEIP), we introduce \textit{XMatch} and \textit{YMatch} as the matching position of \textit{XQuery} and the matching position of \textit{YQuery} for the query rule \textit{qr}. These two variables indicate the current position of the item that to be matched in \textit{XQuery} or \textit{YQuery}. If \textit{XMatch} or \textit{YMatch} is equal to the size of \textit{XQuery} or \textit{YQuery}, then it means that \textit{XQuery} or \textit{YQuery} has been matched exactly. For a sequential rule \textit{tr} = \{$X$\} $\rightarrow$ \{$Y$\}, if \textit{XMatch} is equal to the size of \textit{XQuery} and \textit{YMatch} is equal to the size of \textit{YQuery}, at this time, this rule satisfies \textit{XQuery} $\subseteq$ $X$ and \textit{YQuery} $\subseteq$ $Y$. It is a target sequential rule of the query rule \textit{qr}.
\end{definition}

For example, given a query rule \textit{qr} $=$ \{$a$, $b$\} $\rightarrow$ \{$c$\} and a current sequential rule \{$a$\} $\rightarrow$ \{$c$\}, we can know that \textit{XMatch} and \textit{YMatch} are both equal to 1. Because the size of \textit{YQuery} of query rule \textit{qr} is 1, it means that \textit{YQuery} has been matched exactly. Given another sequential rule \textit{tr} $=$ \{$a$, $b$, $d$\} $\rightarrow$ \{$c$\}, \textit{XMatch} and \textit{YMatch} are equal to 2 and 1 respectively. Since \textit{tr} is exactly matched, \textit{tr} is also one of the target sequential rules of \textit{qr}.

\textbf{Strategy 4: unpromising extended items pruning strategy (UEIP)}. For those rule-growth-based algorithms, expansion is an important operation for them. We use UEIP to avoid invalid expansions. Since we specify that expansions must follow the order $\succ_{lex}$, then when the expandable item is larger than the matching item, this means that related expanded rules can no longer carry the matching item. Therefore, for the left-expansion, we specify that the expanded item cannot be larger than the item located at position \textit{XMatch} of \textit{XQuery}. As for the right-expansion, we specify that the expanded item cannot be larger than the item located at position \textit{YMatch} of \textit{YQuery}. When \textit{YQuery} has not been fully matched, the left-expansion cannot be performed. This is because once the left-expansion is performed, the right-expansion cannot be performed, and if \textit{YQuery} is not matched exactly before the left-expansion is performed, then it is even less likely to be matched exactly afterwards. In addition, as we discuss before, the two variables of rule occurrence can be used to find expandable items, but they are not efficient enough. To further improve the process of finding expandable items, we use two variables, \textit{expandRight} and \textit{expandLeft}. \textit{expandRight} denotes the position where the right-expansion search begins, and \textit{expandLeft} denotes the position where the left-expansion search terminates. Then there will be \textit{expandRight} $=$ \textit{max}(\textit{firstItemset}, \textit{leftEnd}) and \textit{expandLeft} $=$ \textit{min}(\textit{lastItemset}, \textit{rightEnd}).

For example, given a query rule \textit{qr} $=$ \{$a$, $b$\} $\rightarrow$ \{$c$\} and a current sequential rule \{$a$\} $\rightarrow$ \{$c$\}, we can know that an item $d$ can not be left expanded. It is because item $d$ is larger than item $b$ according to order $\succ_{lex}$. Besides, in sequence $s_5$, \textit{expandRight} $=$ \textit{max}(\textit{firstItemset}, \textit{leftEnd}) $=$ \textit{max}(1, 1) $=$ 1 and \textit{expandLeft} $=$ \textit{min}(\textit{lastItemset}, \textit{rightEnd}) $=$ \textit{min}(3, 3) $=$ 3.

\begin{definition}[\textit{CountMap} for optimization]
	\rm We also design two hash tables, \textit{leftCountMap} and \textit{rightCountMap}, to optimize the algorithm. Both \textit{leftCountMap} and \textit{rightCountMap} are two \textit{CountMaps} that record the items that do not appear in the query rule and their corresponding value. As for the \textit{leftCountMap}, the value is the total number of sequences that the first position of the item $e$ in the sequence is less than the \textit{rightEnd} of the sequence. As for the \textit{rightCountMap}, the value is the total number of sequences that the last position of the item $e$ in the sequence is greater than the \textit{leftEnd} of the sequence. If an item $e$ satisfies \textit{leftCountMap}($e$) + \textit{rightCountMap}($e$) $\textless$ \textit{minsup}, then it is also be filtered when using UTP. When using UEIP, if \textit{leftCountMap}($e$) $\textless$ \textit{minsup}, then this item will not be performed a left-expansion. Similarly, if \textit{rightCountMap}($e$) $\textless$ \textit{minsup}, then this item will not be performed a right-expansion.
\end{definition}

For example, in Table \ref{table1}, if \textit{minsup} $=$ 2, $f$ is removed from the database because of \textit{leftCountMap}($f$) + \textit{rightCountMap}($f$) $=$ 1 $\textless$ \textit{minsup}. Given a query rule \textit{qr} $=$ \{$a$, $b$\} $\rightarrow$ \{$c$\} and a current sequential rule \{$a$, $b$\} $\rightarrow$ \{$c$\}, here \{$d$\} can be performed a left-expansion, while can not be performed a right-expansion. It is because \textit{leftCountMap}($d$) $=$ 3 $\ge$ \textit{minsup}, while \textit{rightCountMap}($d$) $=$ 1 $\textless$ \textit{minsup}.

\subsection{Proposed TaSRM Algorithm}

Inspired by RuleGrowth \cite{fournier2015mining} and some ideas of targeted pattern querying, we propose our TaSRM algorithm for targeted sequential rule mining. The core pseudocode of the TaSRM algorithm is given in Algorithm \ref{AL:TaSRM}, Algorithm \ref{AL:RIGHTEXP}, and Algorithm \ref{AL:LEFTEXP} that represent the main, right-expansion, and left-expansion procedures, respectively.

\begin{algorithm}[!htbp]
	\caption{The TaSRM algorithm}
	\label{AL:TaSRM}
	\LinesNumbered

	\KwIn{$\mathcal{D}$; \textit{qr} = \{\textit{XQuery}\} $\rightarrow$ \{\textit{YQuery}\}; \textit{minsup}; \textit{minconf}.} 
	
	\KwOut{all target sequential rules of \textit{qr}.}
	
	initialize \textit{TMPM} $\leftarrow$ $\emptyset$, \textit{leftCountMap} $\leftarrow$ $\emptyset$, \textit{rightCountMap} $\leftarrow$ $\emptyset$\;
	\For{$s$ $\in$ $\mathcal{D}$}{
		\If{$s$ does not contain \textit{XQuery}}{
			remove $s$ from $\mathcal{D}$; \qquad(\textbf{UTP Strategy})
		}
		update \textit{TMPM}\;
	}
	use \textit{TMPM} to remove unpromising items; \qquad(\textbf{UIP Strategy}) \\
	update \textit{leftCountMap} and \textit{rightCountMap}\;
	remove unpromising items and obtain \textit{sids} of all frequent items (\textit{F1})\;
	initialize \textit{firstLeftI} $\leftarrow$ the first item of \textit{XQuery}, \textit{firstRightI} $\leftarrow$ the first item of \textit{YQuery}\;
	\For{$m$ $\in$ \textit{F1}}{
		\For{$n$ $\in$ \textit{F1} \&\& $m$ $\not$= $n$}{
			initialize \textit{pruneIJ} $\leftarrow$ \textit{flase}, \textit{pruneJI} $\leftarrow$ \textit{flase}\;
			\If{$m$ $\textgreater$ \textit{firstLeftI} $\vert\vert$ $n$ $\textgreater$ \textit{rightLeftI} $\vert\vert$ \textit{leftCountMap}($m$) $\textless$ \textit{minsup} $\vert\vert$ \textit{rightCountMap}($n$) $\textless$ \textit{minsup}}{
				\textit{pruneIJ} = \textit{true}; \qquad(\textbf{URP Strategy})
			}
			\If{$n$ $\textgreater$ \textit{firstLeftI} $\vert\vert$ $m$ $\textgreater$ \textit{rightLeftI} $\vert\vert$ \textit{leftCountMap}($n$) $\textless$ \textit{minsup} $\vert\vert$ \textit{rightCountMap}($m$) $\textless$ \textit{minsup}}{
				\textit{pruneJI} = \textit{true}; \qquad(\textbf{URP Strategy})
			}    
			calculate \textit{sids}(\{$m$\} $\rightarrow$ \{$n$\}), \textit{sids}(\{$n$\} $\rightarrow$ \{$m$\})\;
			\If{!\textit{pruneIJ} \&\& $\vert$\textit{sids}(\{$m$\} $\rightarrow$ \{$n$\})$\vert$ $\ge$ \textit{minsup}}{
				calculate \textit{XMatch}, \textit{YMatch}\;
				\If{\textit{XMatch} == \textit{XQuery}.\textit{size}() \&\& \textit{YMatch} == \textit{YQuery}.\textit{size}()}{
					calculate \textit{conf}(\{$m$\} $\rightarrow$ \{$n$\})\;
					\If{\textit{conf}(\{$m$\} $\rightarrow$ \{$n$\}) $\ge$ \textit{minconf}}{
						output \{$m$\} $\rightarrow$ \{$n$\};
					}             
				}
				\ElseIf{\textit{YMatch} == \textit{YQuery}.\textit{size}()}{
					call \textbf{EXPANDLEFT}($r$ $=$ \{$m$\} $\rightarrow$ \{$n$\}, \textit{sids}($m$), \textit{sids}(\{$m$\} $\rightarrow$ \{$n$\}), \textit{XMatch})\;
					call \textbf{EXPANDRIGHT}($r$ $=$ \{$m$\} $\rightarrow$ \{$n$\}, \textit{sids}($m$), \textit{sids}($n$), \textit{sids}(\{$m$\} $\rightarrow$ \{$n$\}), \textit{XMatch}, \textit{YMatch});
				}
				\Else{
					call \textbf{EXPANDRIGHT}($r$ $=$ \{$m$\} $\rightarrow$ \{$n$\}, \textit{sids}($m$), \textit{sids}($n$), \textit{sids}(\{$m$\} $\rightarrow$ \{$n$\}), \textit{XMatch}, \textit{YMatch}); \qquad(\textbf{UEIP Strategy})
				}
			}
			\If{!\textit{pruneJI} \&\& $\vert$\textit{sids}(\{$n$\} $\rightarrow$ \{$m$\})$\vert$ $\ge$ \textit{minsup}}{
				similar to the pseudo-code on lines 23-36...
			}
		}
	}
	
\end{algorithm}

\begin{algorithm}[!htbp]
	\caption{The EXPANDRIGHT procedure}
	\label{AL:RIGHTEXP}
	\LinesNumbered	
	\KwIn{$r$ $=$ \{$I$\} $\rightarrow$ \{$J$\}: the current sequential rule; \textit{sids}($I$): a sid set of all sequences containing the atencedent of $r$; \textit{sids}($J$): a sid set of all sequences containing the consequent of $r$; \textit{sids}($r$): a sid set of all sequences containing $r$; \textit{XMatch}: the matching position of \textit{XQuery}; \textit{YMatch}: the matching position of \textit{YQuery}.}
	initialize \textit{FS} $\leftarrow$ $\emptyset$\; 
	\For{\textit{sid} $\in$ \textit{sids}($r$)}{
		calculate \textit{expandRight}; \qquad(\textbf{UEIP Strategy})\\
		\For{item $e$ in the sequence whose identifier is \textit{sid}, from \textit{expandRight} to the size of the sequence}{
			\If{$e$ $\textgreater$ the item in the \textit{YMatch} position of the sequence}{
				continue; \qquad(\textbf{UEIP Strategy})
			}
			\If{\textit{rightCountMap}($e$) $\textless$ \textit{minsup}}{
				continue; 
			}    
			calculate \textit{sids}(\{$I$\} $\rightarrow$ \{$J$\} $\cup$ $e$)\; 	    
			update \textit{FS}\;
		}
	}
	\For{$e$ $\in$ \textit{FS} \&\& $\vert$\textit{sids}(\{$I$\} $\rightarrow$ \{$J$\} $\cup$ $e$)$\vert$ $\ge$ \textit{minsup}}{
		calculate \textit{newYMatch}\;
		\If{\textit{XMatch} == \textit{XQuery}.\textit{size}() \&\& \textit{newYMatch} == \textit{YQuery}.\textit{size}()}{
			calculate \textit{conf}(\{$I$\} $\rightarrow$ \{$J$\} $\cup$ $e$)\;
			\If{\textit{conf}(\{$I$\} $\rightarrow$ \{$J$\} $\cup$ $e$) $\ge$ \textit{minconf}}{
				output \{$I$\} $\rightarrow$ \{$J$\} $\cup$ $e$;
			}             
		}
		\ElseIf{\textit{newYMatch} == \textit{YQuery}.\textit{size}()}{
			call \textbf{EXPANDLEFT}(\{$I$\} $\rightarrow$ \{$J$\} $\cup$ $e$, \textit{sids}($I$),  \textit{sids}(\{$I$\} $\rightarrow$ \{$J$\} $\cup$ $e$), \textit{XMatch})\;
			call \textbf{EXPANDRIGHT}(\{$I$\} $\rightarrow$ \{$J$\} $\cup$ $e$, \textit{sids}($I$), \textit{sids}($J$ $\cup$ $e$), \textit{sids}(\{$I$\} $\rightarrow$ \{$J$\} $\cup$ $e$), \textit{XMatch}, \textit{newYMatch});
		}
		\Else{
			call \textbf{EXPANDRIGHT}(\{$I$\} $\rightarrow$ \{$J$\} $\cup$ $e$, \textit{sids}($I$), \textit{sids}($J$ $\cup$ $e$), \textit{sids}(\{$I$\} $\rightarrow$ \{$J$\} $\cup$ $e$), \textit{XMatch}, \textit{newYMatch}); \qquad(\textbf{UEIP Strategy})
		}    	
	}
	
\end{algorithm}

\begin{algorithm}[!htbp]
	\caption{The EXPANDLEFT procedure}
	\label{AL:LEFTEXP}
	\LinesNumbered	
	\KwIn{$r$ $=$ \{$I$\} $\rightarrow$ \{$J$\}: the current sequential rule; \textit{sids}($I$): a sid set of all sequences containing the atencedent of $r$; \textit{sids}($r$): a sid set of all sequences containing $r$; \textit{XMatch}: the matching position of \textit{XQuery}.}
	initialize \textit{FS} $\leftarrow$ $\emptyset$\; 
	\For{\textit{sid} $\in$ \textit{sids}($r$)}{
		calculate \textit{expandLeft}; \qquad(\textbf{UEIP Strategy})\\
		\For{item $e$ in the sequence whose identifier is \textit{sid}, from 0 to \textit{expandLeft}}{
			\If{$e$ $\textgreater$ the item in the \textit{XMatch} position of the sequence}{
				continue; \qquad(\textbf{UEIP Strategy})
			}
			\If{\textit{leftCountMap}($e$) $\textless$ \textit{minsup}}{
				continue; 
			}    
			calculate \textit{sids}(\{$I$\} $\cup$ $e$ $\rightarrow$ \{$J$\})\; 	    
			update \textit{FS}\;
		}
	}
	\For{$e$ $\in$ \textit{FS} \&\& $\vert$\textit{sids}(\{$I$\} $\cup$ $e$ $\rightarrow$ \{$J$\})$\vert$ $\ge$ \textit{minsup}}{
		calculate \textit{newXMatch}\;
		\If{\textit{newXMatch} == \textit{XQuery}.\textit{size}()}{
			calculate \textit{conf}(\{$I$\} $\cup$ $e$ $\rightarrow$ \{$J$\})\;
			\If{\textit{conf}(\{$I$\} $\cup$ $e$ $\rightarrow$ \{$J$\}) $\ge$ \textit{minconf}}{
				output \{$I$\} $\cup$ $e$ $\rightarrow$ \{$J$\};
			}             
		}
		\Else{
			call \textbf{EXPANDLEFT}(\{$I$\} $\cup$ $e$ $\rightarrow$ \{$J$\}, \textit{sids}($I$), \textit{sids}(\{$I$\} $\cup$ $e$ $\rightarrow$ \{$J$\}), \textit{newXMatch}); 
		}    	
	}
	
\end{algorithm}

In the main TaSRM algorithm procedure, $\mathcal{D}$, \textit{qr}, \textit{minsup}, and \textit{minconf} are used as input variables, and the algorithm output all target sequential rules of \textit{qr}. TaSRM first initializes three hash tables including \textit{TMPM}, \textit{leftCountMap}, and \textit{rightCountMap} (Line 1). After initialization, TaSRM removes those sequences from the database $\mathcal{D}$ that do not contain the \textit{XQuery} of the query rule \textit{qr} and updates \textit{TMPM} (Lines 2-7). And then, based on UIP, TaSRM uses \textit{TMPM} to remove all unpromising items (Line 8). It also updates \textit{leftCountMap} and \textit{rightCountMap} to further remove unpromising items (Lines 9-10). Meanwhile, the {sids} of all frequent items will be obtained (Line 10). \textit{firstLeftI} and \textit{firstRightI} are used to denote the first item in \textit{XQuery} and \textit{YQuery} (Line 11). Subsequently, TaSRM uses these frequent items to generate all 1 * 1 sequential rules. For any two items $m$ and $n$ in the set of frequent items, TaSRM uses \textit{pruneIJ} and \textit{pruneJI} to determine whether \{$m$\} $\rightarrow$ \{$n$\} and \{$n$\} $\rightarrow$ \{$m$\} can be a qualified 1 * 1 sequential rule (Line 14). Based on URP, \textit{leftCountMap}, and \textit{rightCountMap}, then \textit{pruneIJ} and \textit{pruneJI} get updated to true or false (Lines 15-20). TaSRM calculates two sid sets of sequential rules  \{$m$\} $\rightarrow$ \{$n$\} and \{$n$\} $\rightarrow$ \{$m$\} (Line 21). For the sequential rule \{$m$\} $\rightarrow$ \{$n$\}, TaSRM determines whether this sequential rule satisfies the minimum requirements of \textit{minsup}. If it satisfies, \textit{XMatch} and \textit{YMatch} are computed, and then TaSRM further decide whether it is a qualified target sequential rule of \textit{qr} (Lines 23-29). Finally, according to \textit{XMatch} and \textit{YMatch}, the corresponding left-expansion and right-expansion recursions are performed (Lines 30-36). As for the sequential rule \{$n$\} $\rightarrow$ \{$m$\}, this process is similar (Lines 38-40).

The \textbf{EXPANDRIGHT} procedure is shown in Algorithm \ref{AL:RIGHTEXP}. It first initializes the \textit{FS}, which records the expandable items and corresponding sids set of sequences containing the expanded sequential rules (Line 1). For the sequence containing the current sequential rule $r$, \textbf{EXPANDRIGHT} scans the sequence according to the position \textit{expandRight} to find the expandable items (Lines 3-4). These extendable items need to satisfy the requirements of UEIP and \textit{rightCountMap} (Lines 5-10). If the right-expansion can be performed on the item $e$, then the set of sids for the extended sequential rule \{$m$\} $\rightarrow$ \{$n$\} $\cup$ $e$ is computed, and the \textit{FS} is updated (Lines 11-12). After that, \textbf{EXPANDRIGHT} scans the items in \textit{FS} and calculates \textit{newYMatch} for those items that make $\vert$\textit{sids}(\{$m$\} $\rightarrow$ \{$n$\} $\cup$ $e$)$\vert$ greater than \textit{minsup} (Lines 15-16). If \textit{XQuery} and \textit{YQuery} are matched exactly and the confidence of the new sequential rule is greater than \textit{minconf}, then it is outputted (Lines 17-22). Finally, according to \textit{newYMatch} and \textit{expandRight}, \textbf{EXPANDRIGHT} decides whether to continually perform the left-expansion and right-expansion for the new sequential rule (Lines 23-29).

The \textbf{EXPANDLEFT} procedure is shown in Algorithm \ref{AL:LEFTEXP}. It is basically the same process as Algorithm \ref{AL:RIGHTEXP}. The difference is that it does not need two maintenance matching position variables anymore. When scanning the sequence, the position is from 0 to \textit{expandLeft}. Because of the limitation of space, no more tautology here.

\section{Experiments} \label{sec:experiments}

In this section, to comprehensively evaluate the TaSRM algorithm, three variants of it are designed to better understand the effectiveness of the pruning strategies used. They are TaSRM$_{V1}$, TaSRM$_{V2}$, and TaSRM$_{V3}$. TaSRM$_{V1}$ is the simplest variant of TaSRM, which uses only UTP and a post-processing technique. Compared with TaSRM$_{V1}$, TaSRM$_{V2}$ uses more UIP and URP to discover all target sequential rules. As for TaSRM$_{V3}$, it uses all proposed pruning strategies and an optimization including UTP, UIP, URP, UEIP, and \textit{CountMap} for optimization. TaSRM$_{V3}$ is not required to use a post-processing technique. This is because it uses the matching positions and only the target sequential rules are saved. Because there is no related algorithm about targeted sequential rule mining, we use baseline RuleGrowth \cite{fournier2015mining} as a comparison algorithm to ensure that the number of obtained target sequential rules is correct. The post-processing technique is also adopted to RuleGrowth, and thus it can only save related target sequential rules. We compare the variants of TaSRM with RuleGrowth in terms of the running time, memory consumption, scalability, and different query cases.

We programmed these algorithms in Java language and conducted our experiments on a computer with a 64-bit Win10 operating system, Ryzen 5-3600 CPU, and 8 GB RAM. We make all source code and datasets available at GitHub\footnote{\url{https://github.com/DSI-Lab1/TaSRM}}. The datasets used for the experiments are described in Section \ref{dataset}.

\subsection{Datasets for the Experiment}
\label{dataset}

To better evaluate the performance of the variants of TaSRM on different type of datasets, we selected six datasets for our experiments. Four real datasets and two synthetic datasets are selected. These datasets contain both dense and sparse feature, long and short sequences to allow us to fully evaluate the algorithms. These four real datasets are all single-item-based datasets, which are \textit{BMSWebView1}, \textit{Bible}, \textit{Accidents10k}, and \textit{Sign}, respectively. As for the synthetic datasets, they are \textit{Syn10k} and \textit{Syn20k}, both of which are multiple-items-based datasets and are generated from IBM data generator \cite{agrawal1995mining}. These datasets or their full datasets can be obtained from the open source website SPMF\footnote{\url{http://www.philippe-fournier-viger.com/spmf/}}. Besides, we randomly selected six query rules for these six datasets. The details of all datasets and query rules used in experiments are shown in Table \ref{DATA}, and datasets are described as follows.

$ \bullet $ \textit{\textbf{BMSWebView1}} is a dataset converted from the click data of an e-commerce website. Although there are some long sequences in this dataset, most of the records are short sequences.

$ \bullet $ \textit{\textbf{Bible}} is a book dataset where each sentence in the book Bible is transformed into a sequence. It can be seen as a moderately dense dataset.

$ \bullet $ \textit{\textbf{Accidents10k}} is a 10k dataset of traffic accident data from FIMI. It contains only a few hundred items, but the sequences in it are very long. The original dataset consisted of nearly 340,000 records.

$ \bullet $ \textit{\textbf{Sign}} is a sign language dataset containing 267 items and 730 sequences. At the same time, it is also a very dense dataset.

\begin{table}[!htbp]
	\caption{Details of the different experimental datasets and query rules}  
	\label{DATA}
	\centering
	\resizebox{.95\columnwidth}{!}{
		\begin{tabular}{|c|c|c|c|c|c|c|}	
			\hline
			\textbf{Dataset} & \textbf{$\vert$$\mathcal{D}$$\vert$} & \textbf{$\vert$\textit{I}$\vert$} & \textbf{\textit{AVI}} & \textbf{\textit{AVL}} & \textbf{\textit{Query rule}}\\ \hline   \hline     
			BMSWebView1 & 59601 & 497	& 1.0 & 2.51 &  \{12299, 12343\} $\rightarrow$ \{12355, 12399\} \\ \hline
			Bible & 36369 & 13905	& 1.0 & 21.64 & \{8, 356\} $\rightarrow$ \{10, 46\}  \\ \hline
			Accidents10k &10000	& 310 & 1.0 & 33.92 & \{12, 18\} $\rightarrow$ \{21, 27\}  \\ \hline
			Sign	& 730 & 267	& 1.0 & 27.11 & \{1, 117\} $\rightarrow$ \{143, 253\}  \\ \hline
			Syn10k	& 10000 & 7312	& 4.35 & 26.97 & \{196, 4010\} $\rightarrow$ \{7691, 7823\}  \\ \hline
			Syn20k	& 20000 & 7442	& 4.33 & 26.84 & \{2681, 7011\} $\rightarrow$ \{4171, 5001\} \\ \hline
		\end{tabular}	
	}
\end{table}

In Table \ref{DATA}, $\vert\mathcal{D}\vert$ is the size of the dataset and $\vert$\textit{I}$\vert$ is the number of items in the dataset. \textit{AVI} is used to denote the average number of items contained in each itemset in the dataset, and \textit{AVL} denotes the average length of the sequences in the dataset. The query rules chosen for each dataset are also shown in Table \ref{DATA}. For these datasets, we chose query rules with size 2 * 2 for mining. These query rules are randomly selected from the traditional sequential rule mining results.

\subsection{Efficiency Analysis}

\begin{figure*}
	\centering
	\includegraphics[trim=10 0 0 0,clip,scale=0.089]{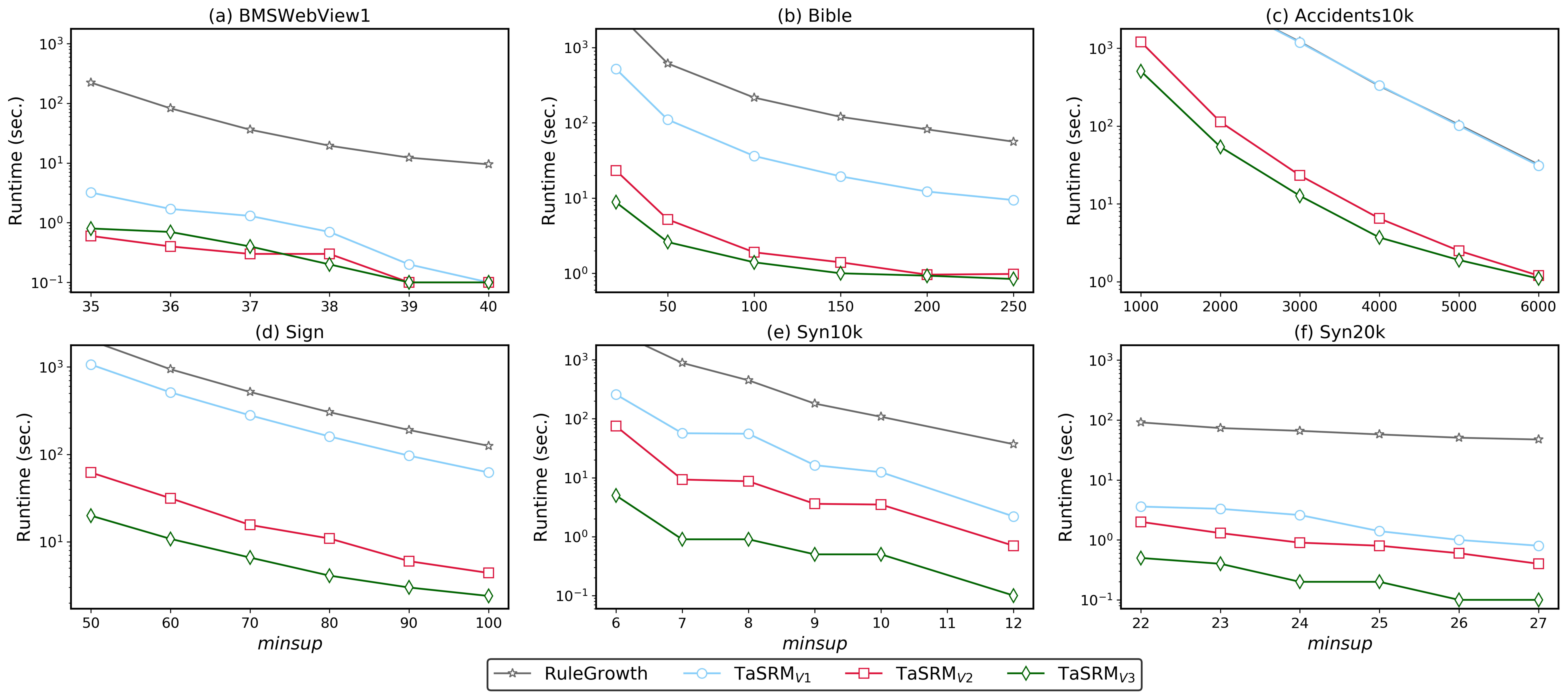}
	\caption{Running time for various \textit{minsup}}
	\label{Runtime}
\end{figure*}

\begin{figure*}
	\centering
	\includegraphics[trim=10 0 0 0,clip,scale=0.089]{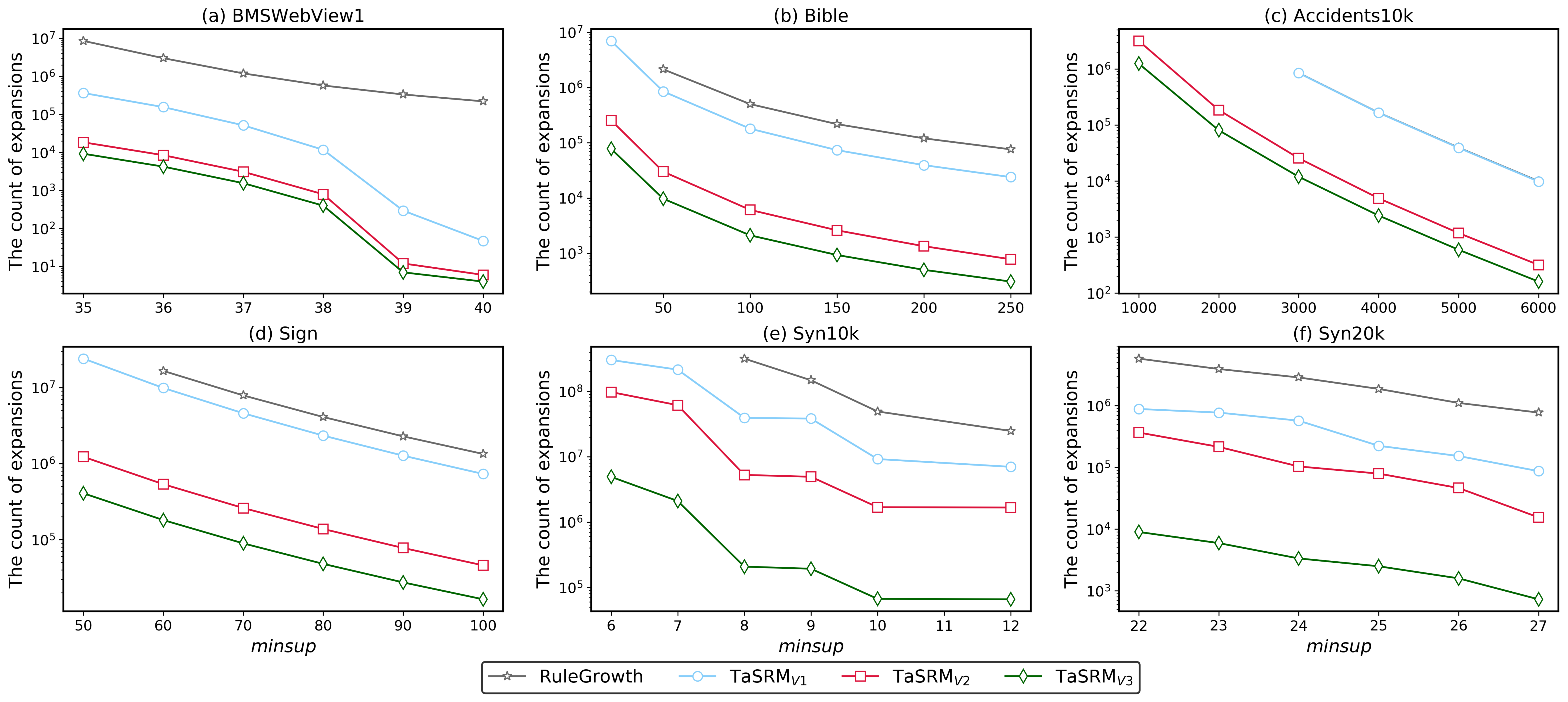}
	\caption{The count of expansions under different \textit{minsup}}
	\label{ExpandCount}
\end{figure*}

In this subsection, we compared the performance of the variants of TaSRM and RuleGrowth in terms of both the running time and the count of expansions. We set the \textit{minconf} to 0.5 and dynamically adjusted the \textit{minsup}. A reasonable \textit{minconf} ensures that the credibility of the target sequential rules are obtained. The experimental results are shown in Fig. \ref{Runtime} and Fig. \ref{ExpandCount}. For those test cases that run longer than 30 minutes, they are not be displayed on the experimental results figure. We use the database filtering rate to indicate what percentage of unrelated sequences are filtered by the UTP. As we discuss above, those unrelated sequences do not affect the generation of target sequential rules. After using UTP, the database filtering rates for these 6 datasets are about 99.93\%, 71.2\%, 0.05\%, 26.4\%, 99.8\%, and 99.8\% respectively. From the results on datasets \textit{Accidents10k} and \textit{Sign}, we can know that the smaller the database filtering rate, the smaller the difference in running time and count of expansions between baseline RuleGrowth and TaSRM$_{V1}$. This is as expected, because the effect of UTP is minor if the dataset contains few unrelated sequences. Once a dataset does not contain any unrelated sequences, then the UTP is not valid. At that time, RuleGrowth and TaSRM$_{V1}$ have almost the same performance.

Overall, as the number of pruning strategies used increases, the variants of TaSRM becomes faster and the count of expansions is reduced as well. The running time and the count of expansions of the variants of TaSRM can be reduced by one or more exponential levels compared to the baseline RuleGrowth on the most datasets. On the \textit{BMSWebView1} dataset, we can see that TaSRM$_{V2}$ is faster than TaSRM$_{V3}$ when the \textit{misup} is lesser than 38. This is because for the efficient enough TaSRM$_{V2}$ and the dataset containing a large amount of short sequences, more sophisticated pruning strategies do not necessarily improve efficiency. On the contrary, additional operations may incur more time overhead. The count of expansions is slightly reduced. Both TaSRM$_{V2}$ and TaSRM$_{V3}$ are able to complete the mining task in less than 1 second. TaSRM$_{V1}$ finishes its task in less than 10 seconds, and RuleGrowth takes tens to hundreds of seconds. As we envision, UTP allows the actually useful sequences to be extracted for mining tasks. On the \textit{Bible} dataset, there is also a small difference between the running time of some test cases in TaSRM$_{V2}$ and TaSRM$_{V3}$. In such test cases, too many pruning strategies sometimes do not work better. TaSRM$_{V2}$ and TaSRM$_{V3}$ still find out all target sequential rules in a short time. Nonetheless, TaSRM$_{V2}$ and TaSRM$_{V3}$ have a difference in the count of expansions. In addition, it can be observed that TaSRM$_{V2}$ has a large improvement compared to TaSRM$_{V1}$. UIP and URP can reduce many unrelated items and invalid 1 * 1 sequential rules on long sequence datasets, thus improving efficiency. Notice that due to the inefficiency of RuleGrowth, when \textit{minsup} is set to 50, it is unable to get any result. On the \textit{Accidents10k} dataset, the running time and the count of expansions of TaSRM$_{V1}$ and RuleGrowth are overlapping. Only 0.05\% unrelated sequences are removed from the dataset, and thus UTP plays a very limited role. Once \textit{minsup} is set to 3000 and below, none of TaSRM$_{V1}$ and RuleGrowth can obtain target sequential rules anymore. On this dataset, the gap between TaSRM$_{V2}$ and TaSRM$_{V3}$ is still not much bigger, but it can be seen that TaSRM$_{V3}$ always achieves a better performance. 

On the dense \textit{Sign} dataset, in terms of running time and the count of expansions, TaSRM$_{V2}$ and TaSRM$_{V3}$ have a significant difference. UEIP and \textit{CountMap} for optimization can work better because there are many long sequences in the dataset \textit{Sign}, allowing the target sequential rules to grow longer. On the \textit{Syn10k} dataset, this optimization effect still exists. Besides, the running time of the variants of TaSRM do not change much when the \textit{minsup} increases from 7 to 8 or from 9 to 10. This is because the number of target sequential rules does not change much, which makes the running time of the variants of TaSRM stable. Furthermore, when \textit{minsup} is set to 8 and below, RuleGrowth cannot generate target sequential rules at a reasonable time. As for the last dataset \textit{Syn20k} we use, each algorithm seems to have little variation in running time, but combined with the observation of the count of expansions, we can know that this is due to little change in the number of target sequential rules. In short, the variants of TaSRM all perform well regardless of any dataset they process. Among all the pruning strategies, the effect of UTP is insufficient on the particular datasets. And none of the other pruning strategies perform too poorly.

\subsection{Memory Evaluation}

\begin{figure*}
	\centering
	\includegraphics[trim=0 0 0 0,clip,scale=0.087]{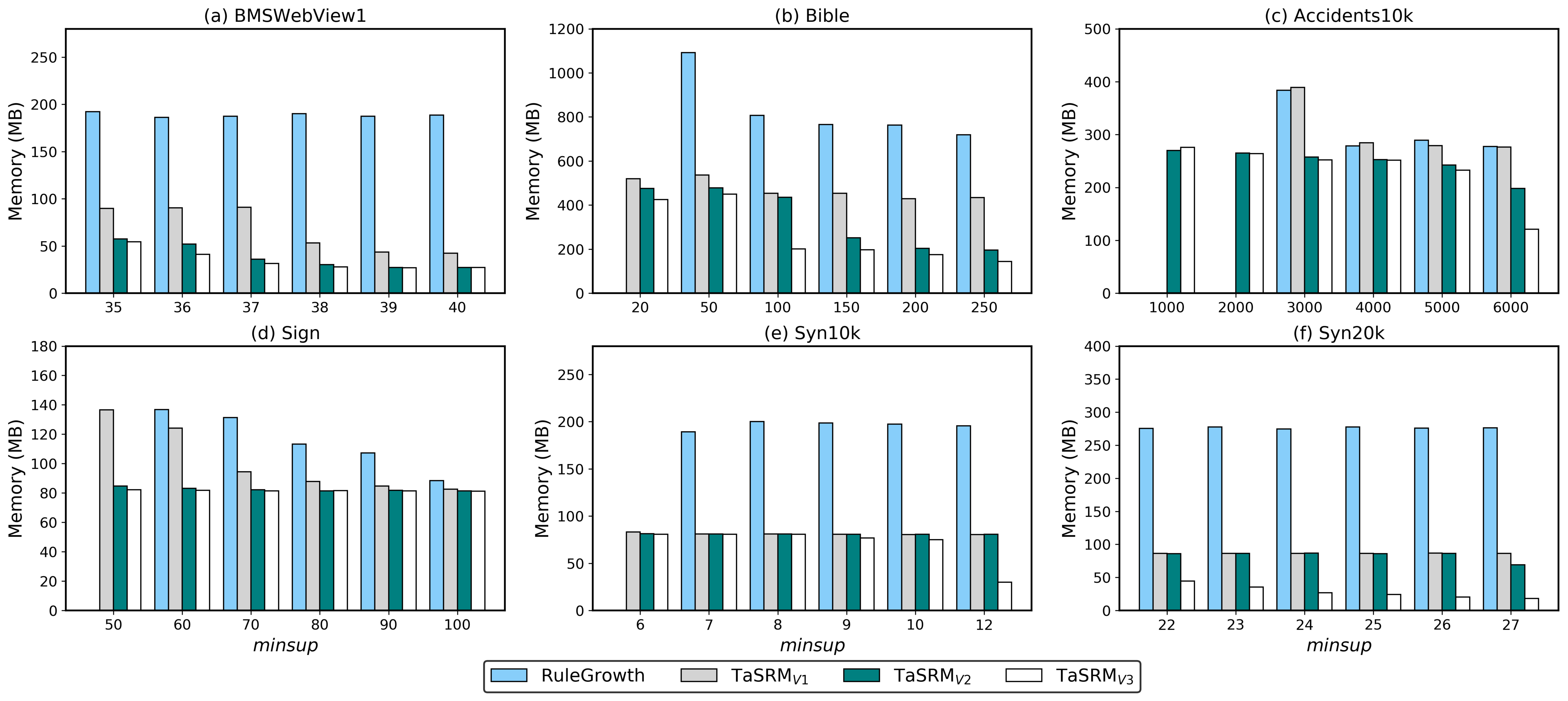}
	\caption{Memory usage for various \textit{minsup}}
	\label{memory}
\end{figure*}

Experimental results on memory usage are shown in Fig. \ref{memory}. Test cases running for more than 30 minutes are not shown memory consumption. We can clearly see that baseline RuleGrowth consume the most memory, especially on datasets \textit{BMSWebView1}, \textit{Syn10k}, and \textit{Syn20k}. On these datasets, there exist many unrelated sequences for the query rules. These unrelated sequences only increase the unnecessary memory consumption and have no effect on mining results. Almost all the test cases on these 6 datasets, TaSRM$_{V3}$ consumes the least memory and TaSRM$_{V2}$ consumes the second least amount of memory. Apparently, the optimization of the two hash tables, \textit{leftCountMap} and \textit{rightCountMap}, does not consume too much memory. 

On the \textit{BMSWebView1}, RuleGrowth uses 1 to 4 times more memory and its memory usage is stable. This is because the range of \textit{minsup} adjustment is not large. The variants of TaSRM exhibit a linear increasing relationship, and the memory usage of TaSRM$_{V2}$ and TaSRM$_{V3}$ are basically the same. This is because TaSRM$_{V3}$ with the more pruning strategies used does not improve the memory optimization much compared to TaSRM$_{V2}$. This is also verified by the performance in running time. On the \textit{Bible} dataset, we can see that when \textit{minsup} is set to 50, RuleGrowth occurs a sudden increase in memory usage. In addition, it also uses twice as much memory compared to TaSRM$_{V1}$. The effect of UTP is still evident. As \textit{minsup} decreases, the gap of memory usage between all variants of TaSRM gradually become close. On the \textit{Accidents10k} dataset, UTP cannot filter many sequences and thus cannot reduce memory consumption. TaSRM$_{V2}$ and TaSRM$_{V3}$ also work well and do not cost much memory. When \textit{minsup} is set to more than 4000, the pruning strategies they use keep their memory consumption stable.

Furthermore, we can see that TaSRM$_{V2}$ and TaSRM$_{V3} $ consume almost the same amount of memory on the \textit{Sign} dataset, staying around 80 MB. This is because the \textit{Sign} dataset contains only a few hundred sequences. RuleGrowth and TaSRM$_{V1}$ both show a significant increasing trend for smaller \textit{minsup} values, while TaSRM$_{V2}$ and TaSRM$_{V3}$ remain stable. When \textit{minsup} is set to 50, TaSRM$_{V1}$ costs almost 140 MB. Note that the variants of TaSRM consumes very little memory in both dataset \textit{Syn10k} and dataset \textit{Syn20k}. There is also not much memory change in these algorithms due to the \textit{minsup} settings with a smaller range. Besides, numerous unrelated sequences are filtered by UTP, avoiding the unnecessary memory consumption. Particularly, TaSRM$_{V3}$ is very memory efficient compared to RuleGrowth on the dataset \textit{Syn20k}. It saves up to ten times more memory used by RuleGrowth. Generally speaking, UTP can indeed reduce a significant memory usage. With the UTP in effect, other pruning strategies can only improve slightly. On some datasets, despite the less effectiveness of UTP, other pruning strategies still work as well as they should. 

\subsection{Scalability}
\label{Sub:scalability}

\begin{figure*}
	\centering
	\includegraphics[trim=0 0 0 0,clip,scale=0.086]{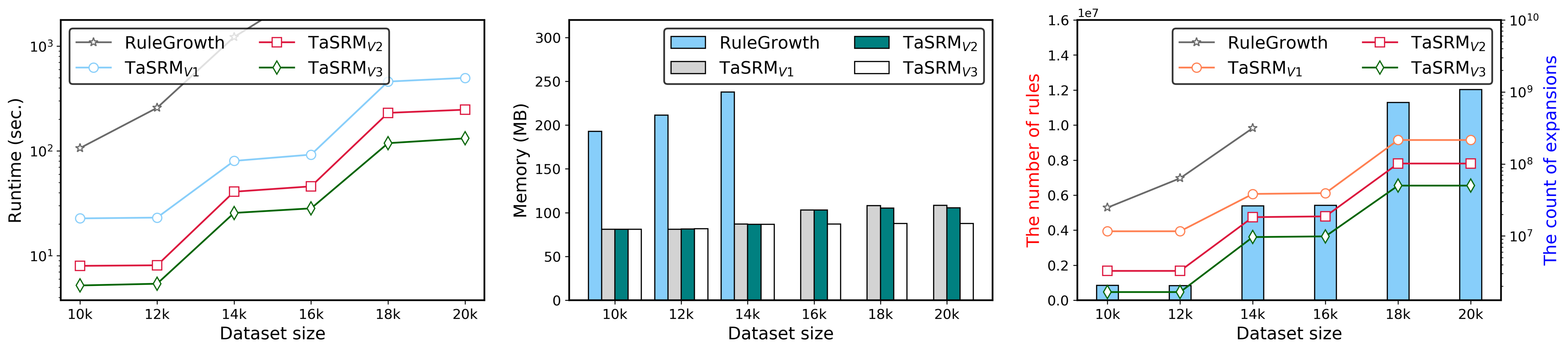}
	\caption{Scalability of the compared approaches when \textit{minsup} = 8 and \textit{minconf} = 0.5}
	\label{scalability}
\end{figure*}

In this subsection, we choose six synthetic datasets for the scalability evaluation of each algorithm, and the dataset size is increased from 10k to 20k. \textit{minconf} is set to 0.5, \textit{minsup} is set to 8, and the query rule is \textit{qr} $=$ \{196\} $\rightarrow$ \{4010\}. The experimental results are shown in Fig. \ref{scalability}. The running time, memory usage, the number of target rules and the count of expansions of all algorithms are presented in three subplots in Fig. \ref{scalability} respectively. In the third subplot of Fig. \ref{scalability}, we show the number of target rules in bars and the count of expansions in dashes. The results are exactly as expected, and we can learn that TaSRM$_{V3}$, which combines all pruning strategies, has the fastest running time and the least memory consumption. Compared to baseline RuleGrowth, the variants of TaSRM can get different degrees of improvement and can even improve hundreds of times. Since the variants of TaSRM all are rule-growth-based algorithms, they have a similar running time increasing trend. When the size of the dataset is greater than 14k, baseline RuleGrowth starts to fail to discover the target rules. The running time and count of expansions of the variants of TaSRM do not change much when the size of the dataset increases from 10k to 12k or 14k to 16k. According to the observation, this is caused by the small change in the number of target sequential rules. We choose a very small \textit{minsup} range for experiments, and thus a minor turning down may make the mining results not change much.

In terms of memory usage, the consumption slowly increases as the size of the dataset increases. The memory usage of RuleGrowth exhibits a linear relationship, and it uses more than twice as much memory compared to TaSRM$_{V1}$. As we expect, the variants of TaSRM consume about the same amount of memory if the UTP is very effective. Their differences are not observed until the size of datasets exceeds 16k. Overall, their consumption is much slower than RuleGrowth. The count of expansions is a good way to verify the growth of the running time for each algorithm. It has a clear trend and is almost identical to the running time. We can make the following conclusions. The variants of TaSRM can exhibit better scalability and the mining efficiency is affected by the number of target sequential rules rather than just the dataset size. As the size of the dataset increases, their running time and memory consumption do not increase as fast compared to that of RuleGrowth.

\subsection{Target Sequential Rule Analysis}

\begin{figure*}
	\centering
	\includegraphics[trim=0 0 0 0,clip,scale=0.087]{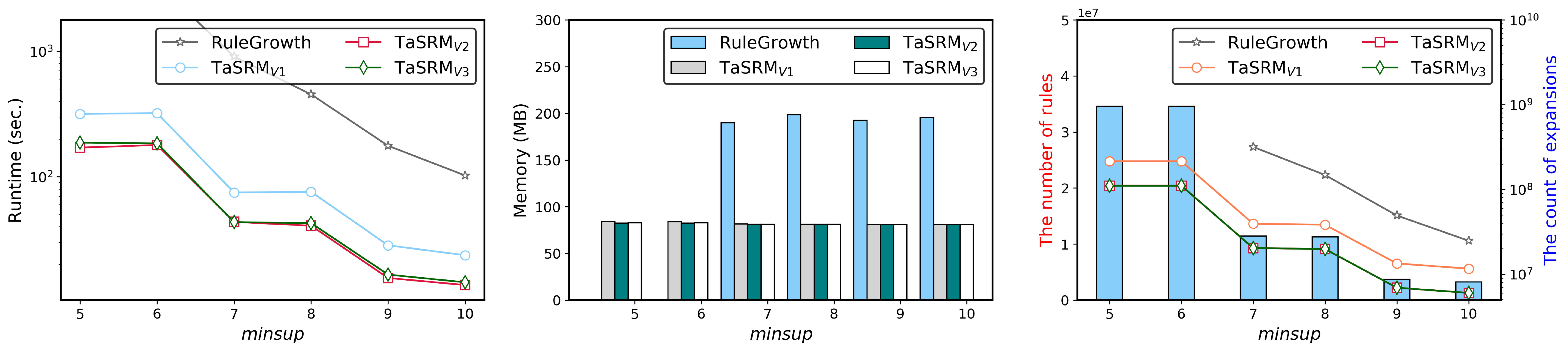}
	\caption{The query case when \textit{minsup} = 8, \textit{minconf} = 0.5, and \textit{qr}$_{left}$ = \{196\} $\rightarrow$ \{\}}
	\label{LeftQuery}
\end{figure*}

\begin{figure*}
	\centering
	\includegraphics[trim=0 0 0 0,clip,scale=0.087]{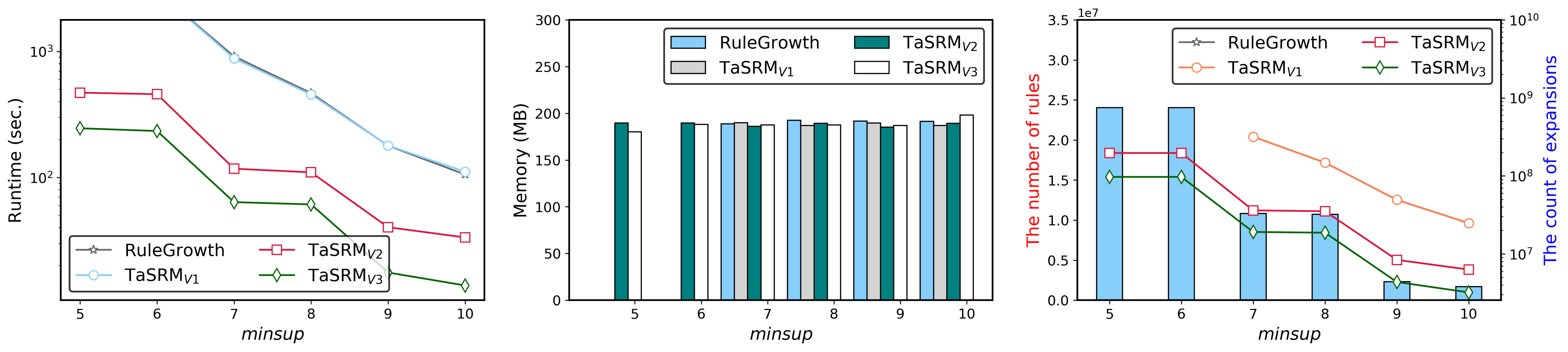}
	\caption{The query case when \textit{minsup} = 8, \textit{minconf} = 0.5, and \textit{qr}$_{right}$ = \{\} $\rightarrow$ \{4010\}}
	\label{RightQuery}
\end{figure*}

In real scenarios, user queries are very diverse. As we discuss in Definition \ref{definition:TSR}, targeted sequential rule mining can be divided into three queries. In this subsection, we evaluate the performance of the algorithms for two special rule query cases. The first query case is a query rule with only antecedent, and the second is a query rule with only consequent. Experimental results are shown in Fig. \ref{LeftQuery} and Fig. \ref{RightQuery}, where the subplots in them show the same aspects as described in Subsection \ref{Sub:scalability}. These two query rules are \textit{qr}$_{left}$ = \{196\} $\rightarrow$ \{\} and \textit{qr}$_{right}$ = \{\} $\rightarrow$ \{4010\} respectively, and the corresponding database filtering rates are 99.03\% and 0\%. According to the filtering rate, we can infer whether UTP is working or not, and the results can verify it.

In the first query case, the variants of TaSRM are still very efficient. It also can be observed that the running time of TaSRM$_{V2}$ is faster than that of TaSRM$_{V3}$. This is because additional pruning strategies and optimizations used in TaSRM$_{V3}$ do not have any effect, and the count of expansions is the same for both algorithms. When the number of target sequential rules does not increase much, the running time and the count of expansions of the variants of TaSRM also change little, while those for RuleGrowth always increase rapidly as \textit{minsup} decreases. The UTP is in effect so that memory usage is also as expected. RuleGrowth cost much memory compared to the variants of TaSRM. In the second query case, the efficiency of baseline RuleGrowth and TaSRM$_{V1}$ is almost the same because the database filtering rate is 0\%, which means that UTP has no effect. Besides, note that there are parts of the running time curves that do not vary much, which can be fully explained by the number of target sequential rules. Unlike the first query case, TaSRM$_{V3}$ has a significant improvement in running time and count of expansions compared to TaSRM$_{V2}$. In terms of memory consumption and for the variants of TaSRM, the second query case uses much more memory than the first query case. This is because UTP has no effect and the adjustment of \textit{minsup} is not great. From the results for the two special query cases, we can conclude that TaSRM$_{V2}$ and TaSRM$_{V3}$ are still able to maintain its efficiency. Since TaSRM$_{V1}$ only uses UTP, it is not suitable for the second query case.

\section{Conclusion and Future Work}   
\label{sec:conclusion}

The advent of the data age has led to a wealth of information being contained in data. Targeted mining is able to extract the most interesting knowledge from the data for users, thus reducing unnecessary resource consumption. Since there is no previous work on targeted sequential rule queries, in this paper, we first define and formulate the problem of targeted sequential rule mining. Possible query situations and the mining results are considered. Meanwhile, we propose an efficient algorithm, namely TaSRM, to obtain all qualified target sequential rules for a query rule. Several pruning strategies and an optimization are proposed to further improve efficiency. Finally, extensive experimental results on different datasets and under different query cases are analyzed to demonstrate the efficiency and effectiveness of TaSRM. We also perform the analysis on special query cases. 

Targeted rule mining is an interesting and novel research direction, and we will explore more related research topics in the future. We can propose dynamic algorithms for the problem that the TaSRM algorithm can only work on the static data. Alternatively, we can introduce some metrics about the targeted sequential rule mining to evaluate the importance of the mining rules. Furthermore, targeted rule mining under big data and noisy data are also the insightful topics.


\ifCLASSOPTIONcompsoc
\else
\fi

\section*{Acknowledgment}

This research was supported in part by the National Natural Science Foundation of China (Grant No. 62002136), Natural Science Foundation of Guangdong Province of China (Grant No. 2022A1515011861), Guangzhou Basic and Applied Basic Research Foundation (Grant No. 202102020277).

\ifCLASSOPTIONcaptionsoff
\newpage
\fi

\bibliographystyle{IEEEtran}
\bibliography{TaSRM.bib}

\end{document}